\documentclass[preprint,3p]{elsarticle}
\usepackage{mymacros}
\usepackage{amssymb}
\usepackage{amsthm}
\usepackage[mathlines]{lineno}
\usepackage{graphicx}
\usepackage{units}
\usepackage{url}
\usepackage{amsmath}
\usepackage{amsfonts}
\usepackage{bm}
\usepackage{textcomp}
\usepackage{subfigure}
\usepackage{multicol}
\usepackage{multirow}
\usepackage{verbatim}
\usepackage{rotating}
\usepackage[colorlinks,linkcolor=red,citecolor=red]{hyperref}
\usepackage{float}
\usepackage{epsfig}
\usepackage{dcolumn}
\usepackage{bm}
\usepackage{color}
\usepackage{pstricks}
\usepackage{pst-node}
\usepackage{times}
\usepackage{indentfirst}
\usepackage[english]{babel}

\biboptions{numbers,sort&compress}

\journal{Physics Letters B}

\begin{document}

\begin{frontmatter}

  \title{{\bf \boldmath Measurement of the absolute branching fraction of
  $\Lambda_c^+\to p K^0_{\mathrm{S}}\eta$ decays}}

  \author{
\begin{small}
\begin{center}
M.~Ablikim$^{1}$, M.~N.~Achasov$^{10,c}$, P.~Adlarson$^{64}$, S. ~Ahmed$^{15}$, M.~Albrecht$^{4}$, R.~Aliberti$^{28}$, A.~Amoroso$^{63A,63C}$, Q.~An$^{60,48}$, ~Anita$^{21}$, X.~H.~Bai$^{54}$
% ...
\\
\vspace{0.2cm}
(BESIII Collaboration)\\
\vspace{0.2cm} {\it
$^{1}$ Institute of High Energy Physics, Beijing 100049, People's Republic of China\\
% ...
$^{4}$ Bochum Ruhr-University, D-44780 Bochum, Germany\\
% ...
$^{10}$ G.I. Budker Institute of Nuclear Physics SB RAS (BINP), Novosibirsk 630090, Russia\\
% ...
$^{15}$ Helmholtz Institute Mainz, Johann-Joachim-Becher-Weg 45, D-55099 Mainz, Germany\\
% ...
$^{21}$ Indian Institute of Technology Madras, Chennai 600036, India\\
% ...
$^{28}$ Johannes Gutenberg University of Mainz, Johann-Joachim-Becher-Weg 45, D-55099 Mainz, Germany\\
% ...
$^{48}$ State Key Laboratory of Particle Detection and Electronics, Beijing 100049, Hefei 230026, People's Republic of China\\
% ...
$^{54}$ University of Jinan, Jinan 250022, People's Republic of China\\
% ...
$^{60}$ University of Science and Technology of China, Hefei 230026, People's Republic of China\\
% ...
$^{63}$ (A)University of Turin, I-10125, Turin, Italy; (B)University of Eastern Piedmont, I-15121, Alessandria, Italy; (C)INFN, I-10125, Turin, Italy\\
$^{64}$ Uppsala University, Box 516, SE-75120 Uppsala, Sweden\\
% ...
$^{c}$ Also at the Novosibirsk State University, Novosibirsk, 630090, Russia\\
% ...
}\end{center}

\vspace{0.4cm}
\end{small}
}

  \begin{abstract}
    Based on 586 $\rm{pb^{-1}}$ of $e^+e^-$ annihilation data collected at a
    center-of-mass energy of $\sqrt{s}=4.6~\rm{GeV}$ with the BESIII detector at
    the BEPCII collider, the absolute branching fraction of $\Lambda_c^+ \to
    p K^0_{\mathrm{S}}\eta$ decays is measured for the first time to be
    $\mathcal{B}(\Lambda_c^+ \to p K^0_{\mathrm{S}}\eta) = (0.414 \pm 0.084 \pm
    0.028)\%$, where the first uncertainty is statistical and the second is
    systematic. The result is compatible with a previous CLEO result on the
    relative branching fraction $\frac{\mathcal{B}(\Lambda_c^+ \to
    p K^0_{\mathrm{S}}\eta)}{\mathcal{B}(\Lambda_c^+ \to p K^-\pi^+)}$, and
    consistent with theoretical predictions of SU(3) flavor symmetry.
    \\
    \\
    \text{Keywords:~~charmed baryon, $\Lambda_c^+$ decays, absolute branching fraction, BESIII}
  \end{abstract}

\end{frontmatter}


\begin{multicols}{2}
\section{Introduction}

The lightest charmed baryon, $\lambdacp$, was first observed in the
1970s~\cite{Knapp:1976qw,Abrams:1979iu}, but only in the past few years has
there been significant progress in precision measurements of its absolute decay
rates~\cite{Asner:2008nq,Ablikim:2019hff}. Driven by these improvements, the
theoretical calculations of charmed baryon decays have greatly improved their
treatment of non-perturbative
effects~\cite{Cheng:2015rra,Lu:2016ogy,Cheng:2018hwl,Geng:2018rse}. However,
more experimental and theoretical efforts are needed to further improve the
understanding of non-factorizable effects, which are non-trivial in the charmed
baryon sector~\cite{Cheng:2018hwl,Zou:2019kzq}. More experimental data are also
desired to test existing theoretical predictions~\cite{Gronau:2018vei}.

The hadronic decay $\lambdacp \to p\kshort\eta$ offers important information
about the dynamics of charmed baryon decays~\cite{Geng:2018upx}. The
charge-conjugate mode is always implied throughout this Letter. The branching
fraction (BF) $\mathcal{B}(\lambdacp \to p\kshort\eta) = \mathcal{B}(\lambdacp
\to p\kzerobar\eta)/2 = (0.79 \pm 0.13~(\rm{stat.}) \pm 0.13~(\rm{sys.}))\%$ has
only been measured relative to $\lambdacp \to pK^-\pi^+$~\cite{Ammar:1995je,pdg}.
However, recent comprehensive theoretical analyses of different BFs for
$\lambdacp$ decays to three-body final states, which assume SU(3) flavor
symmetry, predict that $\mathcal{B}(\lambdacp \to p\kshort\eta)$ is in the range
0.35\% to 0.45\%~\cite{Geng:2018upx,Cen:2019ims}. There is a discrepancy of
approximately 2 standard deviations between the existing experimental result and
those new theoretical calculations. Further, $\lambdacp \to p\kshort\eta$ is
regarded as an important channel to understand the properties of the
intermediate $N(1535)$ resonance~\cite{Xie:2017erh,Pavao:2018wdf}. The $N(1535)$
resonance is puzzling because it decays to final states that contain strange
hadrons, like $\eta N$ and $K\Lambda$~\cite{pdg,Xie:2017erh,Pavao:2018wdf}, but
it does not contain any $\bar{s}s$ component according to the naive constituent
quark model. Thus, further measurements of the $\lambdacp \to p\kshort\eta$
decays may help to verify SU(3) flavor symmetry and may reveal the intermediate
states contributing to these decays.

In this Letter, we present the first absolute measurement of the BF for
$\lambdacp \to p\kshort\eta$ decays based on an $e^+e^-$ collision data sample
corresponding to an integrated luminosity of 586 $\rm{pb^{-1}}$, collected at
the center-of-mass energy of $\sqrt{s}=4.6~\rm{GeV}$ with the BESIII
detector~\cite{Ablikim:2009aa} at the Beijing Electron Positron Collider
(BEPCII)~\cite{Yu:IPAC2016-TUYA01}. At this energy, the $\lambdacp$ baryon is
always produced along with a $\lambdacm$, and the energy remaining does not
allow the production of any additional hadrons. Besides, the previous
analysis~\cite{Ablikim:2015flg}, based on the double tag method and the
simultaneous fit along the 12 $\lambdacp$ decay channels, has reported the total
number of $\lambdacp\lambdacm$ pairs in the same data sample, which gives a
chance to meansure the absolute BF of the other $\lambdacp$ decay channels. In
this analysis, we use the single tag (ST) method in which the $\lambdacp \to
p\kshort\eta$ is reconstructed while the $\lambdacm$ is not.

\section{BESIII Detector and Monte Carlo Simulation}

The BESIII detector is a magnetic spectrometer~\cite{Ablikim:2009aa} located at
BEPCII~\cite{Yu:IPAC2016-TUYA01}. The cylindrical core of the BESIII detector
consists of a helium-based multilayer drift chamber (MDC), a plastic
scintillator time-of-flight system (TOF), and a CsI(Tl) electromagnetic
calorimeter (EMC), which are all enclosed in a superconducting solenoidal magnet
providing a 1.0~T magnetic field. The solenoid is supported by an octagonal
flux-return yoke with resistive-plate-counter muon-identifier modules
interleaved with steel. The acceptance of charged particles and photons is 93\%
of the $4\pi$ solid angle. The charged-particle momentum resolution at $1~{\rm
GeV}/c$ is $0.5\%$, and the $dE/dx$ resolution is $6\%$ for electrons from
Bhabha scattering. The EMC measures photon energies with a resolution of $2.5\%$
($5\%$) at $1$~GeV in the barrel (end cap) region. The time resolution of the
TOF barrel part is 68~ps, while that of the end cap part is 110~ps.

Samples of simulated events produced with the {\sc geant4}-based~\cite{geant4}
Monte Carlo (MC) package, which includes the geometric description of the BESIII
detector and the detector response, are used to determine the detection
efficiency and estimate the backgrounds. The simulation includes the beam energy
spread and initial state radiation (ISR) in the $e^+e^-$ annihilations modeled
with the generator {\sc kkmc}~\cite{ref:kkmc}. Two types of MC samples are
generated in this analysis. The signal MC samples, in which the $\lambdacp$
decays non-resonantly to $\lambdacp\to p\kshort\eta$ while the $\lambdacm$
decays into all possible channels, are used to determine the detection
efficiency. The inclusive MC sample, which consists of the production of
$\lambdacp\lambdacm$ pairs and open-charm mesons, the ISR production of vector
charmonium(-like) states, and the continuum processes incorporated in {\sc
kkmc}~\cite{ref:kkmc}, is used to estimate the backgrounds. Known decay modes
are modeled with {\sc evtgen}~\cite{ref:evtgen} using BFs taken from the
PDG~\cite{pdg}, and the remaining unknown decays from the charmonium states are
modeled with {\sc lundcharm}~\cite{ref:lundcharm}. The final-state radiation
from charged final-state particles is incorporated using the {\sc photos}
package~\cite{photos}.

\section{Event selection}

In this analysis, all the charged tracks are required to satisfy $|\cos\theta| <
0.93$, where $\theta$ is the polar angle of the charged track with respect to
the axis of the MDC. For the proton, the distance of closest approach to the
interaction point (IP) must be within $\pm 10~\rm{cm}$ along the MDC axis
($V_z$) and within $\pm 1~\rm{cm}$ in the plane perpendicular to the MDC axis
($V_r$). The particle identification (PID) algorithm combines the $dE/dx$
information from the MDC and flight time from the TOF to evaluate the likelihood
$\mathcal{L}(h) (h=\pi, K, p)$ for the different final state hadron hypotheses;
to be classified as a proton we require the charged track satisfies both
$\mathcal{L}(p) > \mathcal{L}(\pi)$ and $\mathcal{L}(p) > \mathcal{L}(K)$.

We reconstruct $\kshort\to\pip\pim$ candidates by combining two oppositely
charged tracks. For these tracks, the distance of closest approach to the IP
must be within $\pm 20~\rm{cm}$ along the MDC axis. No requirement on either
$V_r$ or the PID likelihood is applied to maximize the efficiency. The two
tracks are constrained to a common decay vertex by applying a vertex
fit~\cite{Asner:2008nq}, and the corresponding $\chi^2$ is required to be less
than 100. The momenta of the $\pip\pim$ pairs corrected by the vertex fit are
used in the following analysis. Further, the distance of the decay vertex of the
$\kshort$ from the IP is required to be larger than twice the estimated
resolution. The invariant mass of the $\pip\pim$ pairs, $M_{\pip\pim}$, is
required to lie in the mass region $(0.487, 0.511)~\rm{GeV/\textit{c}^2}$.

The $\eta$ candidates are formed by photon pairs with their invariant mass
satisfying the requirement $0.500~\rm{GeV/\textit{c}^2} < M_{\gamma\gamma} <
0.565~\rm{GeV/\textit{c}^2}$. Here, the $\gamma$ candidates are reconstructed
from clusters in the EMC crystals that are not matched with any charged tracks.
The energy deposited in the nearby TOF counters is included in the
photon reconstruction. The deposited energy of any cluster is required to be
larger than 25 MeV in the barrel region ($|\cos\theta| < 0.8$) or 50 MeV in the
end cap region ($0.86 < |\cos\theta| < 0.92$). The measured EMC time of the
cluster, which is defined as the shower time recorded by the EMC relative to the
event start time, is required to be within $(0, 700)~\rm{ns}$ to suppress
electronic noise and energy deposits unrelated to the event. A kinematic fit is
applied to constrain $M_{\gamma\gamma}$ to the nominal mass of the $\eta$
meson~\cite{pdg}. The momenta of the $\gamma\gamma$ pairs corrected by the
kinematic fit are used for further analysis.

The $\lambdacp$ candidates are formed by all combinations of proton, $\kshort$
and $\eta$ candidates in an event. Two kinematic variables, the energy
difference $\Delta E \equiv E_{\lambdacp} - E_{\rm{beam}}$ and the
beam-constrained mass $M_{\rm{BC}} \equiv \sqrt{E^2_{\rm{beam}}/c^4 -
|\vec{p}_{\lambdacp}|^2/c^2}$, are defined to identify $\lambdacp$ candidates.
Here, $E_{\lambdacp}$ and $\vec{p}_{\lambdacp}$ are the reconstructed energy and
momentum of the $\lambdacp$ candidates calculated in the $e^+e^-$ rest frame,
and $E_{\rm{beam}}$ is the average energy of the $e^+$ and $e^-$ beams. All the
candidates are required to satisfy $-0.07~\rm{GeV} < \Delta E < 0.07~\rm{GeV}$
and $2.25~\rm{GeV/\textit{c}^2} < M_{\rm{BC}} <2.3~\rm{GeV/\textit{c}^2}$. If
more than one candidate in an event satisfies all the above requirements, the
best candidate is selected with the following sequence. First, we select
candidates that contain the proton with the largest $\mathcal{L}(p)$. Second, if
there are still multiple candidates, we select those that contain the $\kshort$
with the largest decay length in units of its resolution. Finally, if there are
still multiple candidates after the proton and $\kshort$ requirements, we select
the candidate that contains the $\eta$ with the smallest $\chi^2$ returned by
the kinematic fit.  About 16.9\% of total events in data have multiple
candidates, and the probability to select the correct candidate is 63.1\% based
on a MC study.

\section{Signal yield and BF} 

\begin{figure*}[!tb]
  \centering
   \includegraphics[width=0.40\textwidth]{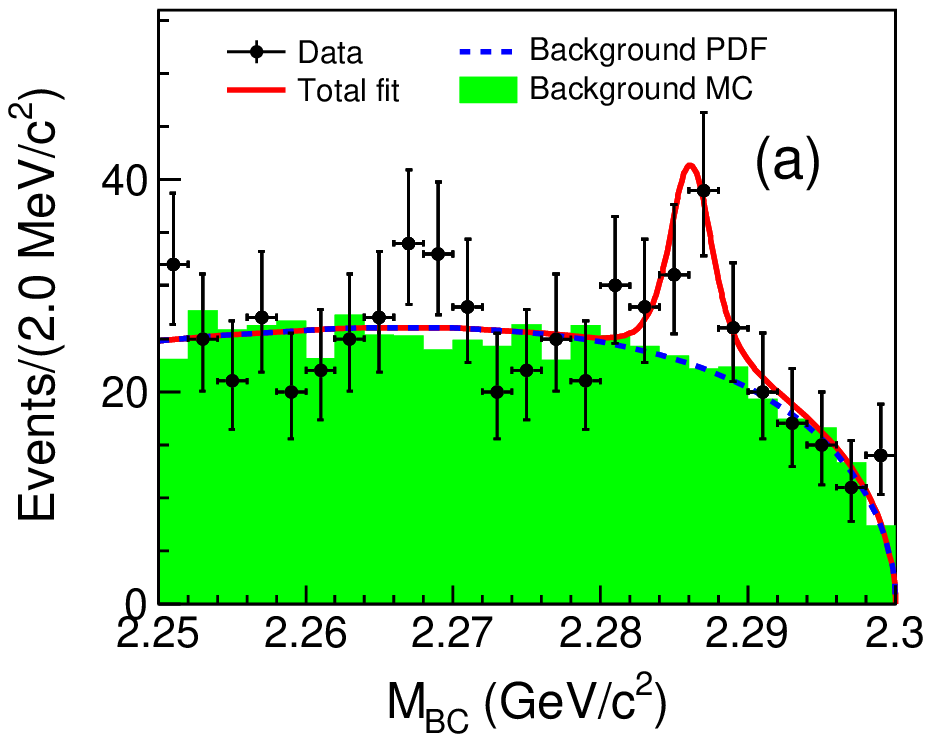}
   \label{fig:1D_mBC} 
   \includegraphics[width=0.40\textwidth]{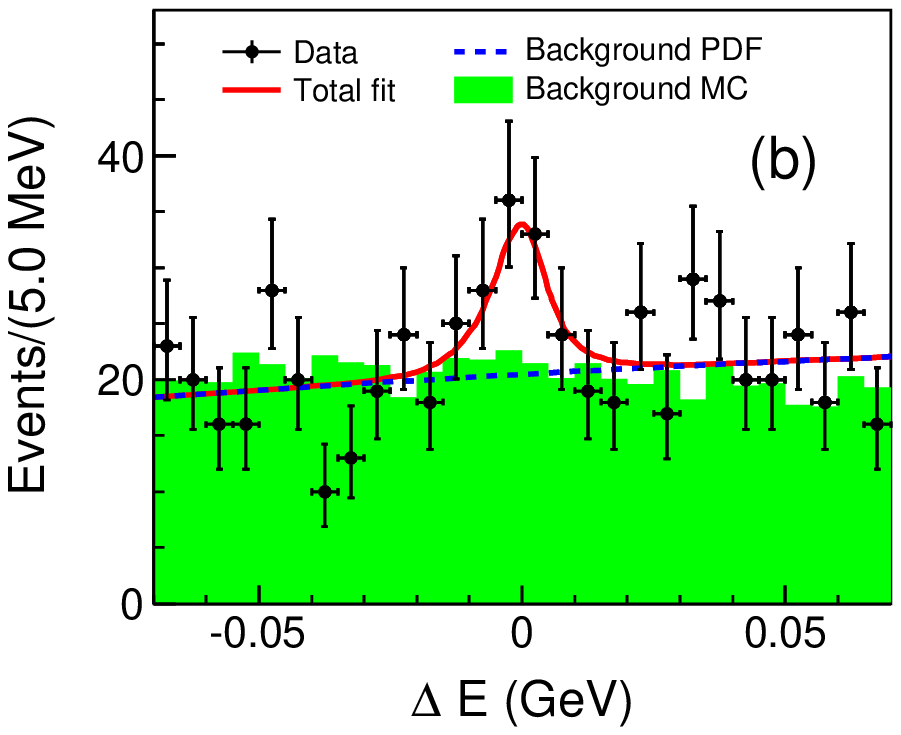}
   \label{fig:1D_deltaE} \\
   \includegraphics[width=0.40\textwidth]{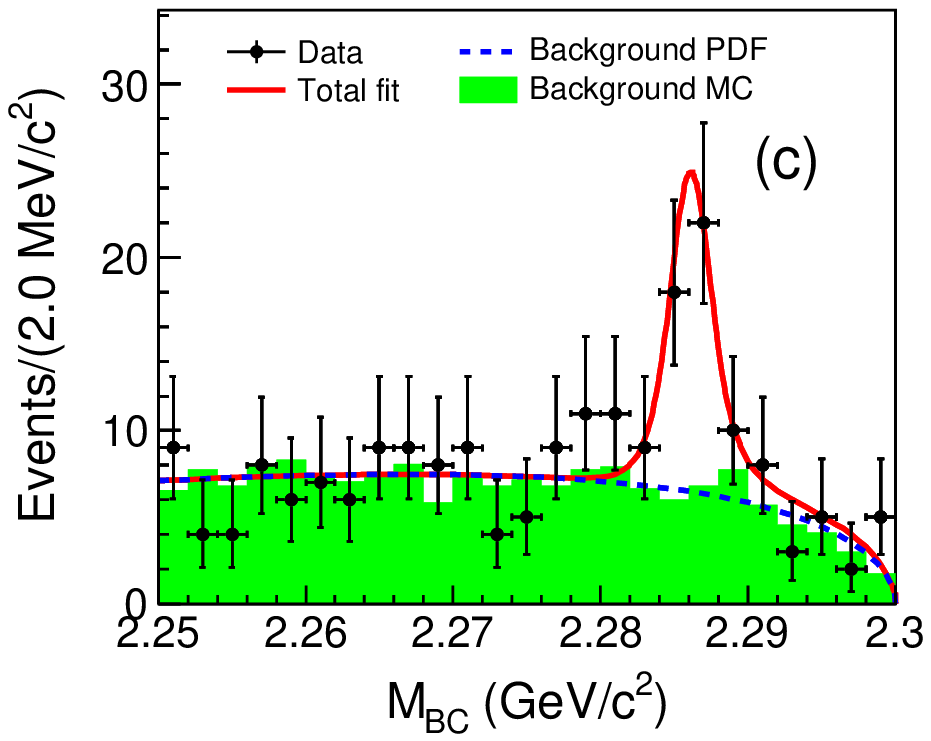}
   \label{fig:1D_mBC_sig}
   \includegraphics[width=0.40\textwidth]{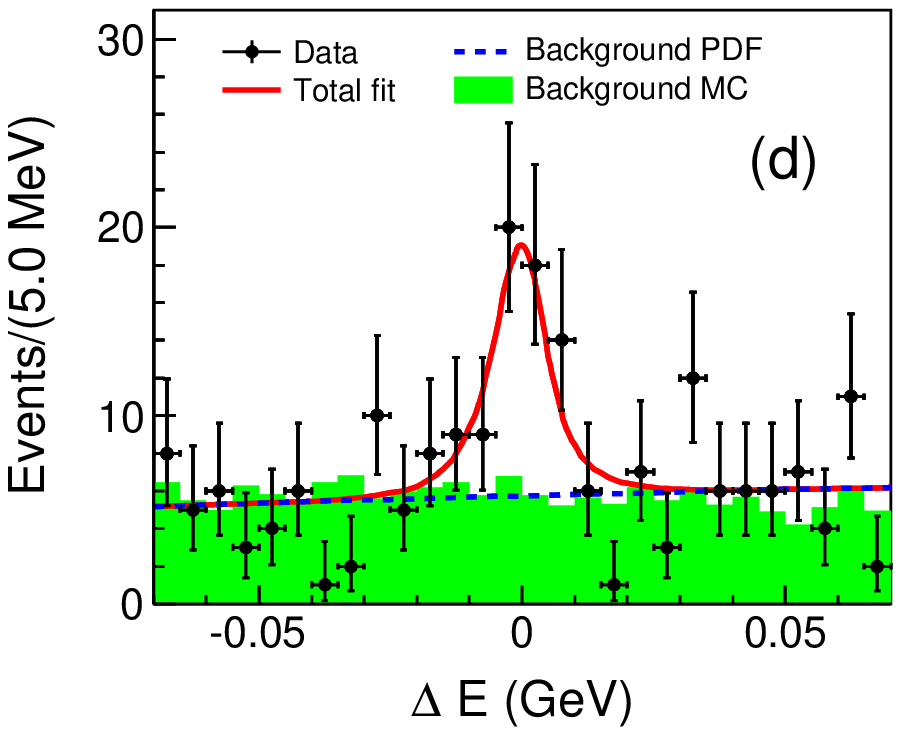}
   \label{fig:1D_deltaE_sig} \caption{One-dimensional projection plots of (a)
   $M_{\rm{BC}}$ and (b) $\Delta E$ distributions, as well as background reduced
   projection plots of (c) $M_{\rm{BC}}$ ($|\Delta E| < 0.02~\rm{GeV}$) and (d)
   $\Delta E$ ($2.28~\rm{GeV/\textit{c}^2} < M_{\rm{BC}} <
   2.295~\rm{GeV/\textit{c}^2}$) with the fit results. Points with error bars
   are data, the solid red line is the total fit result, the blue-dashed line is
   the background contribution, and the green histogram is the background
   distribution from the inclusive MC sample.}
  \label{fig:ST_fit}
\end{figure*}

To determine the signal yield of $\lambdacp \to p\kshort\eta$, an extended
unbinned maximum likelihood fit is performed on the two-dimensional distribution
of $M_{\rm{BC}}$ versus $\Delta E$. Based on the $\kshort$ sideband
($0.02~\rm{GeV/\textit{c}^2}<|m(\pi^+\pi^-)-m(\kshort)|<0.03~\rm{GeV/\textit{c}^2}$)
from data, and inclusive MC studies, no peaking background is found in either
the $M_{\rm{BC}}$ or $\Delta E$ signal region. In the fit, the $M_{\rm{BC}}$ and
$\Delta E$ background shapes are modeled by an ARGUS
function~\cite{Albrecht:1990am} and a second-order Chebyshev polynomial
function, respectively. The end point of the ARGUS function is fixed to the beam
energy $E_{\rm beam} = 2.3~\rm{GeV}$, while all the other
parameters in the background function are free. The $M_{\rm{BC}}$ and $\Delta E$
background shapes are treated as uncorrelated based on MC validation. The signal
shape is the MC-derived two-dimensional shape convolved with Gaussian functions
in each dimension, where the Gaussian functions compensate for the resolution
difference between data and MC samples. The parameters of the Gaussian functions
for $\Delta E$ and $M_{\rm{BC}}$ are fixed to the values determined from the
corresponding one-dimensional fits in data.

Figure~\ref{fig:ST_fit} shows the one-dimensional projection plots of the fit
results. The signal yield is obtained to be $N_{\rm{sig}} = 42.0 \pm 8.5$, where
the uncertainty is statistical only. The statistical significance for the signal
is $7.3\sigma$, which is evaluated by the change of the likelihood between the
nominal fit and the fit without the signal
component~\cite{Asner:2008nq,Narsky:1999kt}. And the significance including the
systematic uncertainties is evaluate to be $5.3\sigma$~\cite{Liu:2015uha}. The
detection efficiency, which is estimated by using the signal MC sample, is
$\varepsilon = (17.57 \pm 0.01)\%$. Thus, the absolute BF of $\lambdacp \to
p\kshort\eta$ can be obtained by
\begin{eqnarray}
  \mathcal{B}(\lambdacp \to p\kshort\eta)=\frac{N_{\rm{sig}}}{2\times
    N_{\lambdacp\lambdacm}\times\mathcal{B}_{\rm{inter}}\times\varepsilon}\quad,
  \label{eq:BF}
\end{eqnarray}
where $\mathcal{B}_{\rm{inter}}=\mathcal{B}(\kshort \to
\pip\pim)\times\mathcal{B}(\eta \to\gamma\gamma)$ is the product of the BFs of
the intermediate states taken from the PDG~\cite{pdg}, $N_{\lambdacp\lambdacm} =
(105.9 \pm 5.3)\times 10^{3}$ is the total number of $\lambdacp\lambdacm$ pairs
in data ~\cite{Ablikim:2015flg}, and the factor 2 is the factor related to the
charge conjugation. Using Eq.~\ref{eq:BF}, we calculate $\mathcal{B}(\lambdacp
\to p\kshort\eta)=(0.414 \pm 0.084)\%$, where the uncertainty is statistical
only.


\section{Systematic uncertainties}

In this work, the systematic uncertainties on the BF measurement include the
uncertainties related to the PID and tracking of the proton, the reconstruction
of the $\kshort$ and $\eta$, the two-dimensional fit, the MC modeling,
$N_{\lambdacp\lambdacm}$ and the BFs of the intermediate states.

The uncertainties related to both the PID and tracking efficiencies of the
proton are estimated to be 1\%, respectively, based on a study of control
samples of $e^+e^- \to p\bar{p}\pip\pim$ and $e^+e^- \to
p\bar{p}\pip\pim\pip\pim$ events in data at $\sqrt{s} >
4.23~\rm{GeV}$~\cite{Ablikim:2015flg}. The uncertainty associated with the
$\kshort$ reconstruction is assigned to be 2.5\% based on studies of control
samples of $J/\psi \to K^*(892)^{\mp}K^{\pm}$ and $J/\psi \to \phi\kshort
K^{\mp}\pi^{\pm}$ decays~~\cite{Ablikim:2015flg}. We use a control sample of
$D^0\to K^-\pip\pi^0$ events to estimate the uncertainty due to the $\eta$
reconstruction~\cite{Ablikim:2017lks,Ablikim:2018byv}. The $\pi^0 \to
\gamma\gamma$ reconstruction efficiencies are obtained in data and MC
simulations, and an $\eta$ reconstruction uncertainty of 1.8\% is assigned by
weighting the efficiency difference to the $\eta$ momentum distribution of the
signal samples.

\begin{figure*}
  \centering
   \includegraphics[width=0.3\textwidth]{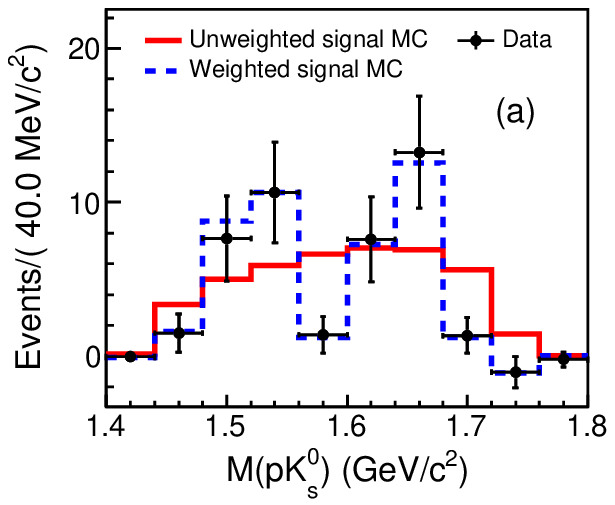} 
   \includegraphics[width=0.3\textwidth]{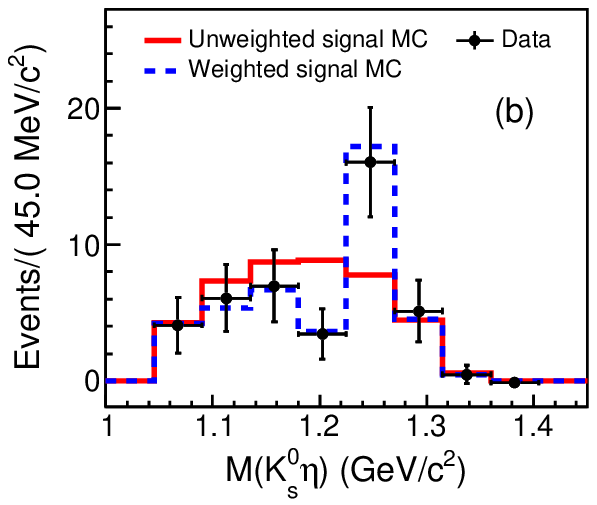} 
   \includegraphics[width=0.3\textwidth]{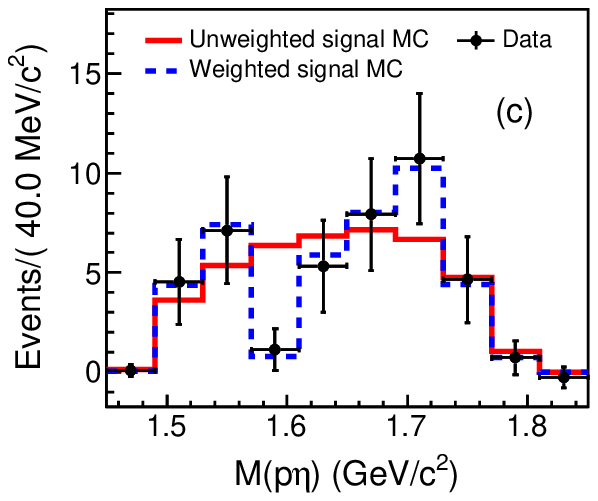} 
  \caption{Distributions of (a) $M_{p\kshort}$, (b) $M_{\kshort\eta}$ and (c)
$M_{p\eta}$ in data and signal MC samples, in which points with error bars are
data after the background subtraction, and the red-solid and blue-dashed
histograms show the unweighted and weighted signal MC samples, respectively.}
  \label{fig:Mass_distributions}
\end{figure*}

The uncertainty associated with the two-dimensional fit consists of the
uncertainties of the signal and background shapes. To estimate the uncertainty,
20000 pseudo-data-sets are sampled with replacement data according to the
bootstrap method~\cite{Efron:1979bxm}. For each pseudo-data-set, a
two-dimensional fit is performed, in which the mean values and standard
deviations of the two Gaussian functions in the signal shape are randomly varied
by $\pm 1\sigma$ around their nominal values, the endpoint of the ARGUS function
is randomly varied by $\pm 0.2\mev$ around its nominal value, and the background
shape of the $\Delta E$ distributions is replaced by a first-order Chebyshev
polynomial function. The pull
of the fit result ${\rm p}(N_{\rm{sig}})$ is defined as
\begin{eqnarray}
  {\rm p}(N_{\rm{sig}}) =
  \frac{N_{\rm{sig}}^{\rm{pseudo}}-{N_{\rm{sig}}^{\rm{real}}}}{\sigma_{N}^{\rm{pseudo}}},
  \label{eq:pull_sys}
\end{eqnarray}
where $N_{\rm{sig}}^{\rm{real}}$ is the fit yield of real data,
$N_{\rm{sig}}^{\rm{pseudo}}$ and $\sigma_{N}^{\rm{pseudo}}$ are the fit yield
and its statistical uncertainty of the pseudo-data, respectively. The
distribution of ${\rm p}(N_{\rm{sig}})$ values is fitted with a Gaussian
function, and the mean value is obtained to be $-0.1364 \pm 0.0074$. The bias in
${\rm p}(N_{\rm{sig}})$ is 13.64\% times the statistical uncertainty, which is
calculated to be 2.8\%, and as no correction for this bias is applied, we use
2.8\% as the systematic uncertainty related with the two-dimensional fit.

The uncertainty related to the MC model of the signal process is evaluated by
reweighting the signal MC sample, which is generated assuming no intermediate
resonances, according to the $M_{p\kshort}$, $M_{\kshort\eta}$ and $M_{p\eta}$
invariant mass distributions observed in data. Here, $M_{ij}$ is the invariant
mass of the combination of particles $i$ and $j$.
Figure~\ref{fig:Mass_distributions} shows the distributions of these invariant
masses in data after the background subtraction using the \mbox{\em sPlot}
method~\cite{Pivk:2004ty} based on the nominal fitting. The detection efficiency
based on the modified MC sample is calculated to be 17.19\%, and the relative
difference to the nominal efficiency, 2.2\%, is taken as the corresponding
systematic uncertainty.

The uncertainty of $N_{\lambdacp\lambdacm}$ in data is reported to be 4.6\% in
Ref.~\cite{Ablikim:2015flg}. The BFs of intermediate states from the
PDG~\cite{pdg} are $\mathcal{B}(\kshort\to\pip\pim) = (69.20 \pm 0.05)\%$ and
$\mathcal{B}(\eta\to\gamma\gamma) = (39.41 \pm 0.20)\%$.  The uncertainty from
these two BFs, which is dominated by the latter one, is 0.5\%.

The total systematic uncertainty, which is determined by adding all sources of
systematic uncertainties in quadrature, is calculated to be 6.8\%. The
systematic uncertainties are summarized in Table~\ref{tab:sys_err}.

\begin{table}[H]
\caption{Systematic uncertainties.}
  \begin{center}
  \begin{tabular}{c|c}
      \hline \hline
			Source & Uncertainty~(\%)\\
			\hline  
			Proton PID & 1.0 \\
			Proton Tracking & 1.0 \\
			$\kshort$ reconstruction & 2.5 \\
			$\eta$ reconstruction & 1.8 \\
			Two-dimensional fit & 2.8 \\
			Signal model  & 2.2 \\
      $N_{\lambdacp\lambdacm}$ & 4.6 \\
			$\mathcal{B}_{\rm inter}$ & 0.5 \\
			\hline
      Total & 6.8 \\
      \hline\hline
    \end{tabular}
    \label{tab:sys_err}
  \end{center}
\end{table}

\section{Summary}

Based on 586 $\rm{pb^{-1}}$ of $\e^+e^-$ collision data collected at $\sqrt{s} =
4.6~\rm{GeV}$ with the BESIII detector and the BEPCII collider, the absolute BF
for $\lambdacp \to p\kshort\eta$ is determined for the first time.  The BF is
measured to be $\mathcal{B}(\lambdacp \to p\kshort\eta) = (0.414 \pm 0.084 \pm
0.028)\%$, where the first uncertainty is statistical and the second one is
systematic. The statistical significance of this measurement is larger than
7$\sigma$, and the significance becomes 5.3$\sigma$ when both statistical and
systematic uncertainties are taken into account. This result is compatible with
the CLEO result $\mathcal{B}(\lambdacp \to p\kshort\eta) = (0.8 \pm
0.2)\%$~\cite{Ammar:1995je} within $1.5\sigma$. Our result is consistent with
theoretical predictions based on SU(3) symmetry~\cite{Geng:2018upx,Cen:2019ims}.
With present statistics, it is not practical to measure any intermediate $N^*$
states, as suggested in Refs.~\cite{Xie:2017erh,Pavao:2018wdf}.

\section{Acknowledgments}
The BESIII collaboration thanks the staff of BEPCII and the IHEP computing
center for their strong support. This work is supported in part by National Key
Basic Research Program of China under Contract No. 2020YFA0406400,
2020YFA0406300; National Natural Science Foundation of China (NSFC) under
Contracts Nos. 11425524, 11575193 11625523, 11635010, 11735014, 11822506,
11835012, 11935015, 11935016, 11935018, 11961141012,11805086, U1732266; the
Chinese Academy of Sciences (CAS) Large-Scale Scientific Facility Program; the
Youth Innovation Promotion Association of CAS under Contract No. 2016153; Joint
Large-Scale Scientific Facility Funds of the NSFC and CAS under Contracts Nos.
U1732263, U1832207; CAS Key Research Program of Frontier Sciences under
Contracts Nos. QYZDJ-SSW-SLH003, QYZDB-SSW-SLH039, QYZDJ-SSW-SLH040; 100 Talents
Program of CAS; INPAC and Shanghai Key Laboratory for Particle Physics and
Cosmology; Fundamental Research Funds for the Central Universities of China; ERC
under Contract No. 758462; German Research Foundation DFG under Contracts Nos.
443159800, Collaborative Research Center CRC 1044, FOR 2359, GRK 214; Istituto
Nazionale di Fisica Nucleare, Italy; Ministry of Development of Turkey under
Contract No. DPT2006K-120470; National Science and Technology fund; Olle
Engkvist Foundation under Contract No. 200-0605; STFC (United Kingdom); The Knut
and Alice Wallenberg Foundation (Sweden) under Contract No. 2016.0157; The Royal
Society, UK under Contracts Nos. DH140054, DH160214; The Swedish Research
Council; U. S. Department of Energy under Contracts Nos. DE-FG02-05ER41374,
DE-SC-0012069.


\end{multicols}
\end{document}